# Understanding Divergent Thermal Conductivity in Single Polythiophene Chains using Modal Analysis and Sonification

Wei Lv[1], Michael Winters[3], Gil Weinberg[3], Asegun Henry[*,1,2]

[1]George W. Woodruff School of Mechanical Engineering, Georgia Institute of Technology, Atlanta, GA 30332, USA.

[2]School of Materials Science and Engineering, Georgia Institute of Technology, Atlanta, GA, 30332, USA.

[3]The Georgia Institute of Technology Center for Music Technology, Atlanta, GA, 30332, USA.

[*]Corresponding author: ase@gatech.edu

**Abstract**:

We used molecular dynamics simulations, the Green-Kubo Modal Analysis (GKMA) method and sonification to study the modal contributions to thermal conductivity in individual polythiophene chains. The simulations suggest that it is possible to achieve divergent thermal conductivity and the GKMA method allowed for exact pinpointing of the modes responsible for the anomalous behavior. The analysis showed that transverse vibrations in the plane of the aromatic rings at low frequencies ~ 0.05 THz are primarily responsible. Further investigation showed that the divergence arises from persistent correlation between the three lowest frequency modes on chains. Sonification of the mode heat fluxes revealed regions where the heat flux amplitude yields a somewhat sinusoidal envelope with a long period similar to the relaxation time. This characteristic in the divergent mode heat fluxes gives rise to the overall thermal conductivity divergence, which strongly supports earlier hypotheses that attribute the divergence to correlated phonon-phonon scattering/interactions.

**Main text:**

In all substances, energy/heat is transferred through atomic motions. In electrically conductive materials, electrons can become the primary heat carriers, but in all phases of matter, atomic motions are always present and contribute to the thermal conductivity of every object. Here, we use the term "object" instead of "material" to emphasize that thermal conductivity is highly dependent on the actual structure of an object, that is, its internal atomic level structure and its larger nanoscale or microscale geometry [1–6]. In studying the physics of thermal conductivity, tremendous progress has been made over the last 20 years toward understanding various mechanisms that allow one to reduce thermal conductivity in solids [1,7]. This reduction is generally measured relative to a theoretical upper limiting maximum value that arises solely from the intrinsic anharmonicity within the interactions between atoms.

In the limiting case, where the atomic interactions are perfectly harmonic, thermal conductivity tends to infinity because the normal modes of vibration do not interact, thus yielding effectively infinite phonon mean free paths and thermal conductivity. As a result, the notion of anharmonicity is critical, and perfectly harmonic interactions between atoms are physically unreasonable. This is because an asymmetry in the energetic well between atoms is an inherent consequence of finite bonding energy (e.g., at some point, the potential energy surface must become concave down). Although one can approach the harmonic limit at cryogenic temperatures, the notion that divergent thermal conductivity (e.g., anomalous thermal conductivity) might be realizable in a real material at room temperature seems theoretically impossible. However, in 1955 Fermi, Pasta, and Ulam (FPU) [8] made a *"shocking little discovery"* that even an anharmonic system can exhibit infinite thermal conductivity.

FPU conducted a numerical experiment with a one-dimensional chain of anharmonic oscillators by introducing, albeit small, cubic terms to a harmonic potential. The expectation was that any degree of anharmonicity would cause interactions between the normal modes of vibration, leading to attenuation and an eventual equipartition of the mode energy, which would confirm that the system behaves ergodically. The resulting effect of the interactions between modes would then translate to finite thermal conductivity. This, however, was not the case. FPU observed that even though a small amount of anharmonicity did result in mode-mode interactions, it did not lead to equipartition and the system trajectory repeated, indicating non-ergodic behavior. Prior to their discovery, it was thought that the only way to approach infinite thermal conductivity was to approach purely harmonic behavior. Their seminal discovery, however, identified a second potential route towards theoretical realization of divergent thermal conductivity.

Since FPU's discovery six decades ago, many studies have tried to understand the underlying physics of how, why and under what conditions the phenomena FPU discovered will persist [9–14]. As a result, most studies focused on "toy models" that use harmonic interactions with small contributions from higher-order terms. This was done largely to examine how small the higher order terms (e.g., cubic, quartic) must be in order to preserve the phenomenon. In addition, the many questions it spawned have been the subject of experts in non-linear dynamics for decades and tremendous progress has been made [15–17]. Excellent reviews of that progress are available elsewhere[14,18–20] and one of the prevailing theories for understanding the origin of the phenomenon is that of mode-coupling theory, which was pioneered by Lepri, Livi, and Politi [19], and the notion that the reduced dimensionality is the origin of the anomalous thermal transport [19]. Previous studies on FPU problem have been focused on predicting the trends of the divergence of thermal conductivity $\kappa \propto L^n$.[20,21] Mode-coupling theory predicts that in such 1D systems, one will observe long tails in the heat-flux correlation function and extremely slow decay towards equipartition [22]. This theory offers a satisfactory explanation for many of the simpler models that have been studied previously. However, one would intuitively expect that at room temperature, all real materials will exhibit anharmonicity that is far too strong to allow non-ergodic behavior to persist. Nonetheless, Henry and Chen[23] showed that the theoretical phenomenon of divergent thermal conductivity can actually persist in a model of a real material at room temperature. Furthermore, Pereira and Donadio[24] reported that 2D graphene may also have divergent thermal conductivity when subjected to a uniaxial strain. More recently, Cepellotti et al.[25,26] and Lee et al.[27] have identified the collective hydrodynamic phonon transport in graphene and a few other 2D materials. They used density-functional perturbation theory and the solution of the Boltzmann transport equation to calculate the scattering rates between phonon modes in order to study the thermal conductivity of 2D systems. Here, we have relied on a different paradigm for understanding the phonon transport. Instead of a scattering based paradigm, we leverage the Green-Kubo formalism, which is inherently correlation based, instead of scattering based. This ultimately leads to a different interpretation of the observed phenomena, but it is important to acknowledge that there may exist a significant connection between the two views, although rigorously describing that connection is beyond the scope of the present investigation.

**Divergent Thermal Conductivity in Polymer Chains:**

The class of materials that most closely resemble the 1D chains that have been studied previously [28–32] are polymers, specifically individual polymer molecules. Carbon nanotubes were once thought to also possibly exhibit divergence, but calculations performed by Mingo and Broido[29,33] as well as other molecular dynamics (MD) studies[34–37] confirmed that the thermal conductivity is large but still finite (non-divergent). However, for polymer chains, Henry and Chen [23] discovered that even in a realistic model for the interactions between atoms in a polyethylene (PE) chain, it is possible to observe divergent thermal conductivity. The observation entails a direct calculation of the thermal conductivity via equilibrium MD (EMD) simulations, which includes anharmonicity to full order. By using EMD, the thermal conductivity can be calculated from the Green-Kubo formula[38,39], which is derived from linear response theory,

$$\kappa = \frac{V}{k_B T^2} \int_0^\infty \langle Q_z(t) Q_z(t+t') \rangle dt' \qquad (1)$$

In Eq. 1 $\kappa$ is the thermal conductivity along the chain length, $V$ is the system volume, $Q_z$ is the volume averaged heat flux along the chain axis direction (z-direction), and $t$ is time. This approach has been used extensively to study the thermal conductivity of many different classes of materials and has shown good agreement with experiments and first principles based approaches [23,40–43]. The quantity $\langle Q_z(t) Q_z(t+t') \rangle$ in Eq. 1 is the heat flux autocorrelation (HFAC) function. Physically it measures the strength of correlation in the atomic motion. For crystalline solids, where the vibrational modes propagate, the autocorrelation decay effectively measures the time scale over which phonons attenuate, where their energy is dissipated through interactions with other modes. For all materials that have been studied, the function $\langle Q_z(t) Q_z(t+t') \rangle$ decays with increased time delay $t'$ and the integral of the HFAC converges to a finite value. Convergence of Eq. 1 with increasing integration time is usually achieved within 500 ps for all previously studied materials at room temperature, including carbon nanotubes [29,42,44,45]. However, even when extended 10 times longer, to 5 ns, the thermal conductivity of individual PE chains did not converge and continued increasing[23]. This divergence is associated with non-ergodic behavior, as some aspect of the trajectory repeats in order to remain correlated over such a long period of time. However, in the previous work[23,30], the underlying repeating phenomenon was never identified and such a repeating phenomenon cannot be observed with other thermal conductivity calculation methods such as non-equilibrium molecular dynamics (NEMD) simulations [23] as have been utilized by others[46].

Henry and Chen [23] used the EMD Green-Kubo (GK) method to study finite length chains with periodic boundary conditions (PBC). The use of PBC is important as it effectively connects one end of the chain to the other and allows vibrations to propagate indefinitely without ever encountering a boundary. The presumption here is that the chain ends would act as a boundary and would scatter the phonons as they do in regular 3D materials, which is often termed a classical size effect [47]. Allowing modes to propagate indefinitely without ever encountering the chain ends is, in one sense, similar to what would happen in an infinitely long chain. However, the finite size of the simulation domain limits the number of normal modes that can interact and therefore differs from the dynamics of an infinitely long chain in that respect. It is important to emphasize that all EMD simulations employing the GK method, other than that of an individual polymer chain and uniaxial strained graphene, have exhibited convergent results [41,44,48,49].

It is expected that the divergent phenomenon[23] observed for an individual molecule will be lost in a larger structure consisting of multiple/many chains. This is because the intermolecular van der Waals interactions with neighboring chains can disrupt the correlation in each chain leading to finite and convergent thermal conductivity [50]. It is conceivable that boundaries (i.e., the chain ends) or other perturbative stimuli such as phonon-photon scattering could disrupt the persistent correlation and cause finite thermal conductivity. However, it remains to be proven whether or not such perturbative interactions are sufficiently strong to disrupt the non-ergodic behavior in all cases. The main issue is that further study is also needed to understand the origin and mechanism as it is not clear what aspect of the trajectory is repetitious, leading to indefinite correlation. Thus, it is also not clear if the phenomenon observed in polymers differs in any way from the mode-coupling description that describes simpler models.

**Fundamental Questions:**

Towards improving our understanding of this phenomenon, we sought out to answer several fundamental questions that lingered from previous analyses[23,30]:

(1) Is divergent thermal conductivity purely restricted to PE, the simplest polymer? It may be possible that higher levels of heterogeneity in stereochemistry prevent the phenomenon from occurring at all and PE is merely an exceptional case.

(2) Is the phenomenon merely a consequence of the adaptive intermolecular reactive bond order (AIREBO) potential[51] used by Henry and Chen[23]? It may be possible that the behavior previously[23] observed is just a numerical anomaly of the specific potential and parameter set used.

(3) Is the phenomenon caused by the usage of periodic boundary conditions (PBC)? PBC restricts the dynamics to just a finite group of modes, yet allows them to propagate without ever encountering a boundary. This fictitious effect may be the only reason the divergence manifests. One would expect the effects of the boundaries to decay as the chain gets longer, but it would be more conclusive if anomalous behavior somehow manifested without PBC.

(4) What is the actual cause of the phenomenon? Previous theory suggests that the lowest frequency modes' energy correlation decays very slowly[21] and the modes essentially have infinite relaxation times. This explanation is consistent with what would be required if one computed thermal conductivity through the phonon gas model. However, it has not been proven whether or not this explanation explains the divergence observed in a polymer chain.

**Non-Ergodic Behavior:**

A previous study by Henry and Chen[30] suggested that the divergence in PE was caused by cross-correlations between mid-frequency longitudinal acoustic modes. In essence they argued that the normal modes do in fact attenuate, however, the mode-mode interactions remain correlated in time, which violates the *stosszahlansatz* assumption [52] that underlies the Boltzmann transport equation (BTE). The *stosszahlansatz* assumption is an assumption that was made by Boltzmann during the derivation of the Boltzmann transport equation (BTE). Specifically, in order to simplify the particle collision integral, Boltzmann assumed that each collision was independent of all previous collisions and that successive collisions were not correlated. In the context of phonons, this translates to an assumption that successive phonon scattering events are unrelated, and therefore when one collision occurs there is no memory that carries forward or affects subsequent collisions. Henry and Chen [30] introduced a correlation-based paradigm stemming from the GK formula and argued that one cannot use a phonon gas model/BTE (i.e. relaxation time) based formalism to explain the phenomenon, because the *stosszahlansatz* assumption is violated. This was substantiated by estimating the relative contributions of different correlations within the GK formalism, by substituting $Q_z$ with the expression used in the phonon gas model $Q_z = \frac{1}{V}\sum_k \sum_p \hbar \omega v_z \delta n(t)$, where $v_z$ is the group velocity along chain back done direction and $\delta n$ is the deviation from the average occupation.

A key shortcoming of the method they employed was that it could only serve as an estimate [30]. Summing all correlations would not necessarily reproduce the full GK result exactly and thus it is difficult to quantify precisely which interactions (i.e. correlated scattering events) are most responsible. In this study we address the four aforementioned questions, using the following approach: To address question (1) we simulated a different polymer, polythiophene (Pth), which was a polymer of interest for high conductance thermal interface materials [4], that has greater inhomogeneity in stereochemistry (e.g., a more complicated unit cell). To address question (2) we used a different potential, namely the ReaxFF potential [53]. ReaxFF is similar to AIREBO in the sense that it is a bond order potential, however, the mathematical descriptions differ significantly. ReaxFF is constructed to be as general as possible, capturing all conceivable interactions from covalent terms, columbic interactions as well as van der Waals forces. Thus, it serves as a substantial departure from the original AIREBO potential used by Henry and Chen[23]. To address question (3) we used simulations of finite length chains, without PBC, to look for anomalies in the thermal conductivity as a function of chain length $\kappa(L)$. To address question (4) we used a new formalism for calculating the modal contributions to GK thermal conductivity, namely the Green-Kubo modal analysis (GKMA) method[54], since it exactly accounts for each mode's contribution. Using GKMA allows us to study the mode interactions in greater detail with higher fidelity and we also employed sonification techniques[55,56], whereby the heat flux of an individual mode is converted to an audible sound. Sonification[55,56] allows one to identify patterns and features in the data that might go undetected by the eye, since the human ear is much better equipped for time series pattern recognition.

**MD Results:**

We used the ReaxFF potential and confirmed its utility at describing phonon transport in Pth by first computing the thermal conductivity of amorphous Pth as $0.3 \pm 0.1$ Wm$^{-1}$K$^{-1}$, which is in reasonable agreement with experiments $0.19 \pm 0.02$ Wm$^{-1}$K$^{-1}$ [4]. We then proceeded to answer question 1 by conducting MD simulations of chains with PBC to determine whether or not the divergence manifests. The relaxed structure with a 0.796 nm long unit-cell was used as the initial structure. The simulations employ PBC along the axial direction of the chain, used a 0.15 fs time step, and were run at equilibrium in the NVE ensemble for 20 nanoseconds, with 100 picoseconds of equilibration time. EMD simulations of different numbers of unit cells exhibited interesting results for single Pth chains. More than 80 independent simulations in total were conducted to ensure the divergent cases were not anomalies, but instead occurred often enough to be non-negligible in the ensemble average. Although only 20% of the cases diverged, if one considers calculation of the ensemble average, the averaged thermal conductivity will diverge as a result. We calculated the thermal conductivity for 10, 20, 30, 40, 80, 90, 120, and 150 unit cell chains, and the 30, 90 and 150 unit cell length chains exhibited divergent thermal conductivity. Similar to Henry and Chen's [23] results for PE, there were some simulations that exhibited clear convergence and others that exhibit clear divergence, with many cases falling in between these extremes. The only difference between each case was the random initial velocities used. A comparison of example convergent and divergent cases is shown in Fig. 1.

We have carefully examined the energy and momentum conservation as well as the movement of the center of mass during the simulations and have ruled out the possibility that the divergence is due to an aggregated effect of small numerical errors or non-zero bias in the average heat flux. Furthermore, no obvious significant difference in the total heat flux was observed for convergent vs. divergent results. Identification of the true underlying mechanism, however, should show a distinctly different behavior for the convergent and divergent cases. Nonetheless, the results shown in Fig. 1 address questions (1) and (2), namely they show that the divergence in single chains is not only a manifestation of one specific potential or one specific polymer (e.g., PE is not a single special case).

As indicated by question (3), another potentially suspicious cause of the divergence could be the application of PBC to such a 1D structure. To address this question, we tested Pth chains without PBC using one unit cell of rigid atoms (i.e., atoms with infinite mass) on each end. This was done to prevent an individual chain from coiling, which is a potential problem for very long chains. With this modification, phonon propagation becomes restricted by the chain length. These simulations still used EMD but did not employ PBC, used a 0.15 fs time step and were run for 3 ns at equilibrium in the microcanonical ensemble with 100 ps of equilibration time. Five independent simulations were run for each chain length for better averaging. Each data point is averaged over 15 ensembles to avoid statistical fluctuations. The results of the finite chain simulations are shown in Fig. 2 and indicate that a large deviation from the trend at 90 ucs. In the simulations using PBC, we observed divergence for chain lengths at multiples of 30 ucs and, interestingly Fig. 2 shows that even without PBC, we observe anomalous behavior manifests for chains with multiples of 30 ucs. The short length multi-30 ucs chains (30 ucs and 60 ucs) do not behave differently, presumably because when the system length is too small, the long wavelength phonons are constrained by the boundary scattering. However, what is particularly interesting is the difference between 88 ucs and 90 ucs. The two chain lengths only differ by 1.5 nm from the

90 ucs chains, yet the difference in thermal conductivity is a factor of four, which is anomalous. This suggests that the anomalous behavior arises from the specific interactions that can occur in chains with a very specific length, such as multiples of 30 ucs. From Fig. 2, we also observe that the thermal conductivity of 90 ucs and 150 ucs are similar in magnitude. It is likely a very specific relationship and interplay between the mode energies and momenta that gives rise to the anomalous behavior, since slightly shorter or longer chains will have modes with very similar wavelengths and frequencies, but not exactly the same ratios are multiples of 30 ucs. It is also important to note that, because of this, the phenomenon could become less pronounced for longer chains, whereby the presence of other modes opens up additional channels for phonon interaction sequences. This could be the reason the 150 uc chain in Fig. 2, has almost the same thermal conductivity as the 90 uc chain and not a much higher value.

Computing the power spectrum of the HFAC shows that the dominant frequency shifts dramatically between chains with similar lengths and is peaked around 0.1 THz for the anomalous 90 ucs chain. This frequency corresponds to atom vibrations close to 0.05 THz (50 GHz), since $Q$ is proportional to the atom velocities and $Q$ is squared in the HFAC, such that the mode frequency is half that of the HFAC frequency. Figure 3 shows that the heat flux for chains with 90 and 150 vibrating unit cells has a dominant frequency at 0.1 THz, while other chains have different peak frequencies. Figure 3c shows that the modes near 0.05 THz exist in both 88 and 90 ucs chains, yet the frequency peaks in Fig. 3 correspond to significantly different areas of the spectrum.

Upon examination of the HFAC for the chains with PBC and the chains without PBC, approximately 0.8 ns of time delay are needed to reach 120 $Wm^{-1}K^{-1}$. This time scale is three orders of magnitude longer than the time it would take for the modes with frequencies near 0.05 THz to reach the chain ends and scatter. This is important because it suggests that somehow these modes are able to remain correlated even after encountering the chain ends. For comparison, consider the maximum thermal conductivity contribution a mode can have as predicted by the phonon gas model/BTE. For a mode in a 1D system to transfer energy and contribute to the thermal conductivity it must scatter at the chain ends. The group velocity for the modes near 0.05 THz is ~ 820 m/s, which suggests that if the high thermal conductivity is due to modes in this frequency range, and they remain correlated for 3 ns, they are somehow unperturbed by the chain ends since they must encounter the chain ends every 0.0873 ns. Using the phonon gas model and BTE, the thermal conductivity contribution from a mode is given by $\kappa(k,p) = cv_g l$, where $c$ is classical specific heat $k_B/V$, $v_g$ is phonon group velocity, and $l$ the phonon mean free path, which at its maximum, is equal to the chain length. Using this model, the thermal conductivity of this mode can only be as large as 0.0443 $Wm^{-1}K^{-1}$. However, our simulations suggests that these modes must contribute much more in order to reach the finite chain value of ~ 120 $Wm^{-1}K^{-1}$.

In previous studies [19,57], one observes a smooth curve for $\kappa(L)$ and many studies have focused on identifying the exponent in $\kappa \propto L^n$ (n= 1/3 or 2/5). Our results, deviate from this as we see discontinuities in $\kappa(L)$ at multiples of 30 ucs. The fact that there is correspondence between the chain lengths that exhibited the divergence in the PBC cases and the finite chains without PBC that exhibited anomalously high thermal conductivity suggests that the divergence is not caused by the PBC, even though it is enabled by the PBC. Using PBC allows the thermal conductivity to diverge, but the fact that anomalous behavior is also observed at the same chain lengths without PBC, indicates that it is not the cause of the divergence. Instead the divergence is most likely associated with the specific modal interactions in chains with multiples of 30 ucs and not the PBC itself. To try and understand the phenomenon more deeply (e.g., question (4)), we used the recently developed GKMA formalism[54] to study the divergent behavior in more detail. Here, we briefly introduce the GKMA formalism, which is presented in detail elsewhere[54].

**Modal Analysis – Towards Deeper Understanding:**

In the purely harmonic limit, the solutions to the equations of motion are exact and the resulting eigen modes do not interact. Anharmonic systems can be analyzed by using the harmonic modes as a basis set, however, the anharmonicity causes the amplitudes to become time dependent and they cannot be determined analytically. From this perspective, we can still view the vibrations as arising from the modes obtained in the harmonic limit, but the interactions between modes manifest through the time dependent mode amplitudes, which can be studied through a fully anharmonic numerical simulation technique such as molecular dynamics (MD). It is in this sense that the positions and velocities contain all degrees of anharmonicity, and the harmonic eigen modes simply serve as the basis set from which we can interpret/understand them. The GKMA formalism combines the lattice dynamics (LD) formalism with the GK formalism and allows one to determine the thermal conductivity of a system via a direct summation over modal contributions. To accomplish this, we first use the eigen modes determined from LD to write the velocity of every individual atom as,

$$\mathbf{v}(j) = \frac{1}{(Nm_j)^{1/2}} \sum_{\mathbf{k}} \sum_{p} e(j,\mathbf{k},p) \cdot \exp(i \cdot \mathbf{k} \cdot \mathbf{r}(j)) \cdot \dot{X}(\mathbf{k},p) \qquad (2)$$

where

$$\dot{X}(\mathbf{k},p) = \sum_{j} \dot{X}_j(\mathbf{k},p) = \frac{1}{N^{1/2}} \sum_{j} \sqrt{m_j} \exp(-i \cdot \mathbf{k} \cdot \mathbf{r}(j)) \cdot e(j,\mathbf{k},p) \cdot \mathbf{v}(j) \qquad (3)$$

where $j$ denotes the atom, $m_j$ is its mass, $N$ is the number of unit cells in the system, and $e(j,\mathbf{k},p)$ is the polarization vector which gives the direction in which each atom moves. Equations 2 and 3 represent a transformation to and from the normal mode coordinates. Equation 2 gives the modal contributions to every individual atom's velocity. From Eqs. 2 and 3 one can describe the modal contributions to any quantity that is a direct function of the atomic displacements or velocities. To study thermal conductivity, we substituted the modal velocity into the heat flux operator [58], to obtain the modal heat flux

$$\mathbf{Q}(\mathbf{k},p) = \frac{1}{V} \sum_j \left[ E_i \dot{X}_j(\mathbf{k},p) + \sum_j (-\nabla_{r_i} \Phi_j \dot{X}_j(\mathbf{k},p)) \cdot \mathbf{r}_{ij} \right] \qquad (4)$$

We can then take this expression for the heat flux and substitute it directly into the GK expression for modal thermal conductivity,

$$\kappa(\mathbf{k},p) = \frac{V}{k_B T^2} \int_0^\infty \langle \mathbf{Q}(\mathbf{k},p,t)\mathbf{Q}(0)\rangle dt \tag{5}$$

resulting in,

$$\kappa(\mathbf{k},p) = \frac{V}{k_B T^2} \int_0^\infty \left\langle \mathbf{Q}(\mathbf{k},p,t) \cdot \sum_{\mathbf{k}'}\sum_{p'} \mathbf{Q}(\mathbf{k}',p',0) \right\rangle dt \tag{6}$$

These two expressions are the primary results of the GKMA formalism and it has been validated by comparing to experiments as well as other well-established modeling methods [59].

Using Eq. 5, one can compute the correlation of one mode with the entire heat flux at once, or one can substitute the modal contributions to the heat flux in both places to examine the contributions in more detail, through every cross-correlation or only the correlations of interest, via Eq. 6. Since this method is guaranteed to recover the exact value of the total thermal conductivity and the simulations are based on deterministic dynamics, one can repeat simulations of interest using the same initial conditions and examine the contributions in increasing levels of detail gradually, to maximize computational efficiency. With GKMA we can also calculate the eigen mode contributions to thermal conductivity directly and do not have to rely on the phonon gas model/BTE, which becomes questionable in such systems, given the potential violation of the *stosszahlansatz* assumption.

In examining the divergence in Pth, the first objective was to identify which modes are responsible for the divergence by determining which phonon polarization contributes most to the thermal conductivity. Figure 4 shows the contributions from different phonon branches. Figure 4a shows that the transverse acoustic modes that correspond to vibrations in the plane of the aromatic rings (TA-y) and out of the plane (TA-x) contribute most to the total thermal conductivity in 30 unit cell chains. All other branches contribute negligibly to the total thermal conductivity. Since the correlation function between mode heat flux and total heat flux allow having negative value, one would expect some negative contribution temporally if the contribution from the branch is oscillating around zero. However, the total thermal conductivity is always positive. In Fig. 4b we show the divergent and convergent comparison for these two branches. The results are averaged over 35 ensembles (10 divergent cases, 25 convergent cases). If the TA-y and TA-x branches diverge then the total thermal conductivity diverges and vice versa.

Using the GKMA approach, we can also study the correlation between individual branches and individual modes. In the following we focused on the TA-y branch since it contributes most to the thermal conductivity. Using Eq. 6 we examined the TA-y branch's correlation with other branches and as shown in Fig. 4c the autocorrelation of the TA-y branch is most important. With the focus on TA-y mode-mode interactions we computed the individual mode contributions on the TA-y branch to determine which of the modes is most responsible for the divergence. Figure 4d shows that the lowest frequency mode on the TA-y branch contributes the most and when combined with the other 3 lowest frequency modes on the TA-y branch, the three modes (3 out of 30 modes) comprise almost the entire thermal conductivity for the entire branch (Note that this includes the symmetric +k and –k modes, which are identical). The fact that the divergence is associated with the lowest frequency modes seems consistent with the ideas in mode coupling theory. However, mode-coupling theory suggests that the divergence arises because the long – wavelength modes have very slow energy diffusion, in other words, long (infinite) relaxation times.

To determine if infinite relaxation time is the most suitable explanation for the behavior observed in Pth we calculated the relaxation times of these low frequency modes using the method introduced by Ladd *et al.*[60] and McGaughey and Kaviany[61]. By comparing the results for converging and diverging cases using their method, we can determine if the difference in thermal conductivity can be explained by a lack of scattering or a lack of interaction with other modes. The relaxation time of the lowest frequency modes on TA-y branch calculated in divergent and convergent cases are very close (on the order of 100~300ps). Fig.S2 shows the energy autocorrelation functions for a convergent and divergent case in the supplementary material. In both cases the autocorrelations decay. Similar to the phenomenon observed before,[23] the autocorrelations for the divergent cases show a significant resurgence in correlation at much later times, but the integration still remains of similar magnitude to the convergent cases. From the relaxation time analysis, there is no major difference between converging and diverging cases and in both cases the relaxation time is finite (on the order of 100~300 ps). For this reason, the BTE and phonon gas model (PGM) appears insufficient for describing the divergent behavior in Pth, as the only mechanism available in the PGM for attaining infinite thermal conductivity is for a mode to have an infinite relaxation time.

From Fig. 4d it is clear that over the course of 5 ns, the lowest frequency mode on the TA-y branch (labeled TA-$y_1$) contributes 100 $Wm^{-1} K^{-1}$. It should be noted here that the both the positive and negative k-vector contributions are shown summed together (since they are indistinguishable). Such a large contribution from a single mode is anomalous, as 100 $Wm^{-1} K^{-1}$ is higher than most elemental metals (i.e. consisting of many modes and many electrons). However, in Pth, it is remarkable that one single mode has the ability to transfer so much heat. Thus, in Pth the thermal conductivity divergence is primarily dominated by a single special mode. It is important to emphasize that the contribution of this mode is linearly increasing during the 5 ns of correlation time computed from Fig 4d. We have even extended the simulation times by an additional 10 ns (20 ns total yielding, a maximum 10 ns long correlation) and the divergence simply continues along the same path. What is also interesting is that the anomalous conducting mode's frequency is 0.05 THz. This frequency corresponds to the peak in the Fourier transform of $Q$ for the finite chains with lengths that are exact multiples of 30 ucs. For 88 unit cells the peak moves to 0.09 THz, but even on the 88 ucs chain there is a mode on the TA-y branch with a frequency very close to 0.05 THz. It is an anomalous effect of the anharmonicity that two individual chains with such similar lengths (i.e., 88 and 90 ucs) do not exhibit similar modal interactions and thermal conductivity, even though the frequencies and wavelengths in each case are very similar. In addition to the results shown in Fig 4., we have used GKMA to study the correlation of the TA-$y_1$ mode with the other modes on the TA-y branch. The analyses and results are given in the supplementary materials and essentially show that there are strong cross-correlations/interactions amongst the lowest three modes on the TA-y branch that give rise to the divergence.

With the TA-$y_1$ mode now identified and the important interactions with other modes now clarified, the last remaining question was to try and develop a better understanding of the underlying mechanism for the divergence. Visual inspection of the heat flux for the TA-$y_1$ modes did not reveal any obvious differences that would indicate what aspects of their heat flows remain correlated and give rise to the divergent thermal conductivity contributions (see Fig. 5a). However, the integration of the correlation function clearly indicates that something in the heat fluxes of these modes is similar and/or repetitious. It should be noted that persistent correlation need not manifests as a periodic pattern occurring on regular intervals. Instead it could merely be some portion of the signal that is recurring, albeit possibly at irregular intervals (e.g., not necessarily rhythmically). Either scenario or any intermediate between these extremes could result in a divergent integral of the correlation functions. Interestingly, we computed the Fourier transform of the total and mode level heat fluxes for both convergent and divergent cases, as shown in Fig. 5b. Figure 5b shows that there are no major differences in the heat flux spectra in the very low frequency regions. Since there was no obvious difference in the frequency content and no obviously significant features obtained from visual inspection, we sonified the mode level heat fluxes to determine if patterns or any discernable differences would emerge via audible inspection.

Sonification[55] involves slowing down the THz oscillatory data so that it can be studied as an audible sound. The human ear is much better equipped for identifying patterns than the human eye, and thus sonification can be used to identify features within the data that would go undetected when presented visually. Direct sonification of data has been used in other instances for the same reasons and in other situations the data is amended or converted to mimic familiar types of sounds, such as a musical instrument [62,63]. Here, however, we were interested specifically in preserving the original features of the data completely so that the truly distinguishing features between the convergent and divergent cases could be identified and would not be artificially augmented. Attached audio files "file1" and "file2" respectively contain the heat flux of the TA-$y_1$ mode for a clearly convergent and clearly divergent case. The HFAC integral for these two simulations are shown in Fig. s5 in the supplementary materials for reference, which shows that after 5 ns of integration the convergent case's mode thermal conductivity was 8 Wm$^{-1}$ K$^{-1}$ and the divergent case's mode thermal conductivity was 49 Wm$^{-1}$ K$^{-1}$ (6 times larger). In attached audio files 1 and 2, the heat flux was slowed down by a factor of $10^{10}$, such that the 20 ns of MD data plays as a 200 sec sound, which is a shift in time scale that enables some pattern recognition. For example, if the data were slowed down by a factor of $10^8$, the entire MD simulation would play as a short 2 sec sound – too short to distinguish any features. Slowing the data by a factor of $10^{10}$, also causes the low 0.05 THz frequencies of the TA-y modes of interest would correspond to 5 Hz, which is below the 20 Hz threshold of the human ear. However, because the heat flux contains extremely high frequency components, due to interactions with the much higher frequency optical modes that sound like noise, one can hear each oscillation of the mode as a pulsing sound. In this sense, the high frequency noise imparted by the optical modes, which include the motions of the hydrogen atoms, is the only thing that can be heard, but its volume (e.g., loudness) is modulated by the modes low frequency oscillations, which are much larger in amplitude. Inspection of the audio in attached audio files 1 and 2 immediately identified a feature that seemed to be much more prevalent in the divergent case (file 2) than in the convergent case (file 1), namely the repeated instances in the divergent case where the signal slowly fades out and then fades back in.

The time scale in between such events is on the order of 15 seconds, which in the MD simulation corresponds to several hundred picoseconds, close to the mode relaxation time. We believe that the fade in and fade out phenomenon is related to the period between phonon interactions. Conceptually, this supports the hypothesis put forth by Henry and Chen that the divergence cannot be attributed to ballistic phonon transport, but instead is arises from patterned phonon interactions. Individual phonons do in fact experience interactions and scatter, causing the relaxation time to be finite. However, the scattering experience, possibly via the sequence of energy exchanges is similar and repeats in a correlated way. In this sense a thermal perturbation excites a mode and it relaxes, but the mode is later re-excited in a way the mimics the initial excitation and a repetitious pattern of the same de-excitation-excitation cycle persists. It is therefore confirmatory to note that the cycle/correlated feature time scale is the same as the relaxation time, which represents the time scale between excitation and de-excitation events.

In the divergent case these fade in and fade out events occur approximately at the following time marks [15s, 30s, 45s, 58s, 70s]. The difference between the convergent and divergent audio is also easier to hear in audio file3, which combines file1 and file2 into a single stereo file, whereby file1 is played on the left channel and file2 is played on the right channel. One can then hear the periods near [15s, 30s, 45s, 58s, 70s] where the sound fades out on the right ear channel, while the left ear channel plays almost continuously. The identification of this feature was then inspected audibly for a variety of convergent and divergent cases, which qualitatively confirmed that such events happened more often in the divergent cases than the convergent cases.

The purpose of the sonification was to identify a feature that differs between convergent and divergent results, which would then guide the analysis further. The identification of the fade in and out events then guided their visual identification from a plot of the heat flux for the TA-$y_1$ mode. By properly rescaling the time axis so that regions of 400ps to 2000ps are shown, one can then visually observe the features that were more easily identified in the audio files. Figure 5c shows a zoomed in portion of the data in Fig. 5a, where one can see the oscillatory envelope on the TA-y1 heat flux. We then sought out to quantify these large oscillations in the heat flux that generate such large peaks and valleys. The regions of the heat flux where these events occur are similar and repetitious, but occur at irregular intervals. Nonetheless, these features are what repeat throughout the trajectory and they give rise to the persistent correlation. It is not clear if such events are associated with a single scattering event/interaction between modes, as they could also be the result of a series or combination of mode-mode interactions that recur.

To quantify the phenomenon, it was hypothesized that the persistent correlation is a result of these repetitious events, which seemed to be greater in both magnitude and frequency in divergent cases. To test this hypothesis we examined the time history of the TA-$y_1$ heat flux for each individual case and numerically calculate a data series that represents the envelope of the function. This was accomplished by first squaring the mode heat flux so that the result is always positive, as shown in Fig. 5d. The envelope of $Q^2$ was then determined by taking the numerical derivative at every point, via finite difference, and locating points in the function where the function changed from positive to negative slope (a maximum) and negative to positive slope (a minimum). The resulting maximums and minimums were then averaged together over bins in time of 0.0375 ps, which helped to smear/coarse grain the data set, yielding the overall envelope of the function as shown in Fig. 5d. Once the envelope function was determined, we calculated each subsequent maximum and minimum in the envelope $f_{max,i}$, $f_{min,i}$ and calculated their difference to gauge both the size and frequency (150~190) of the peaks and valleys in the envelope,

$$F_{pv} = \sum_{i}(f_{max,i} - f_{min,i}) \qquad (7)$$

where the subscript i denotes nearest neighboring maxima or minima. Since there are many cases in between the extremes of clearly convergent vs. divergent cases as shown in Fig. 1, we then calculated $F_{pv}$ for 20 different independent MD simulations and compared the value of $F_{pv}$ to the value of the thermal conductivity at the end of the 10 ns of correlation time. The result, which is shown in Fig. 6 shows that there is a clear relationship between $F_{pv}$ and thermal conductivity. The greater the magnitude and frequency of the fade in and fade out events, the greater the thermal conductivity, which confirms the hypothesis that these events are the root cause of the divergence in Pth.

With the root cause finally identified we must then examine the corresponding physical interpretation of such a phenomenon. First, it is useful to reason through the meaning of the mode level heat flux. If for example, one were to have a single polymer chain and begin a MD simulation where only one mode is excited, by definition, if one examines all of the mode heat fluxes, one would only see that the excited mode's heat flux is non-zero. This is because the modal contribution to the heat flow is proportional to the mode amplitude. Thus, when only one mode is excited, one would observe the heat flux for the excited mode oscillating at the mode frequency. If all of the atomic interactions were harmonic then one would cease to observe mode-mode interactions and the mode heat flux would simply oscillate at the excited mode frequency indefinitely. However, as anharmonicity is included and mode-mode interactions take place, the frequency content of the excited mode's heat flux will change as it interacts with other modes. Most specifically, as the primary excited mode couples to other modes, one would expect to see some frequency content associated with the interacting modes show up in the primary excited mode's heat flux and vice versa, namely that the mode interacting with the primary excited mode would also exhibit fluctuations with a frequency component associated with the primary excited mode.

With this framework for interpreting the mode level heat flux, it can then be understood that the fade in and fade out events correspond to the mode being excited and then de-exited over the course of several hundred picosecond that is close to the relaxation time of the mode through a particular sequence of interactions that repeat and remain correlated. It is the fact that each of these events occurs in a similar fashion each time that enables the correlation to persist and leads to divergent thermal conductivity. In between such periods, there are substantial interactions with lots of other modes, as indicated by the wide range of frequency content, which can both be seen visually in Fig. 5c and heard audibly as noise in audio files 1-3. These repetitious events are the root cause of the divergence in Pth and they occur to different extents in different independent simulations. To our knowledge the analysis herein serves as one of the most detailed explanations offered to date, for understanding the root cause of anomalous heat conduction in polymers.

We investigated anomalously divergent thermal conductivity (anomalous thermal conductivity) in individual Pth chains, using MD simulations, the GKMA method and sonification. Four important questions regarding the fundamental physics have been answered herein, namely: (1) PE is not a singular/exceptional case of divergent behavior, since it also occurs in a model for Pth. (2) The divergence is not purely a consequence of the AIREBO potential[51] used by Henry and Chen, since it manifested here using the ReaxFF potential. (3) The divergence is not just a numerical anomaly or aggregated error and is not caused by the usage of PBC, since an anomalous increase in thermal conductivity was observed for a finite chain consisting of the an exact multiple of 30 ucs (e.g., the exact chain length required to observe anomalous thermal conductivity when PBC are applied). (4) The origin of the divergence was pinpointed as arising from the lowest frequency modes on the TA-y branch, which clearly exhibits divergent correlations in every instance where the total thermal conductivity diverges. Furthermore, the results showed that the modes responsible for the divergence do in fact have finite relaxation times, and do exhibit significant interactions with other modes e.g., they do scatter and it is not consistent with a ballistic transport interpretation. Lastly, the origin of the divergence was determined through sonification of the mode heat flux, as the lowest frequency mode (50 GHz) exhibited repeated excitation and de-excitation events that repeated more strongly and more frequently in divergent cases. These findings offer new insight into the root causes and underlying physics at play in polymer chains where anomalous heat conduction occurs and thus the analysis herein serves as a significant step forward in our understanding of this intriguing phenomenon that in essence corresponds to anomalous thermal conductivity.

## Acknowledgements

This research was supported Intel grant AGMT DTD 1-15-13 and computational resources were provided by the Partnership for an Advanced Computing Environment (PACE) at the Georgia Institute of Technology and National Science Foundation supported XSEDE resources (Stampede) under grant numbers DMR130105 and TG-PHY130049. We thank Tengfei Luo and Teng Zhang from University of Notre Dame for useful discussion on amorphous Pth and Fang (Cherry) Liu for her assistance with code porting and integration at PACE.

## Contributions

A.H. provided direction and supervised the study. W.L. carried out the simulations and analyzed the data. M.W. performed the sonification analyses, with guidance from G.W. and interactions with W.L. and A.H. W.L., M.W. and A.H. wrote the manuscript. All authors have reviewed, discussed and approved results and conclusions of this article.

## Competing financial interests

The authors declare no competing financial interests.

**Figures**

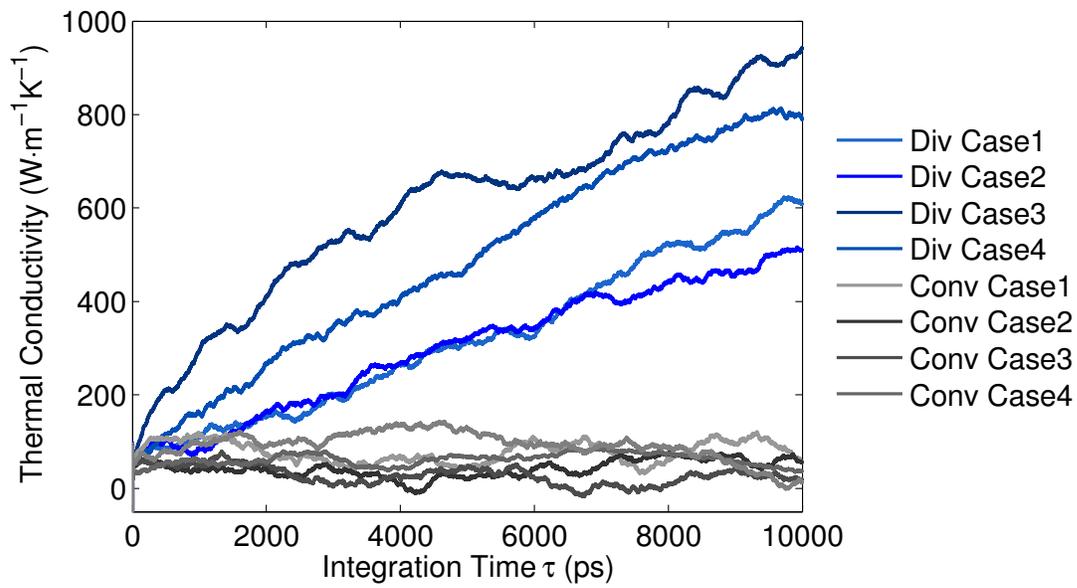

**Figure 1. Green-Kubo thermal conductivity integrals for individual 30 unit cell long polythiophene chains with periodic boundary conditions**

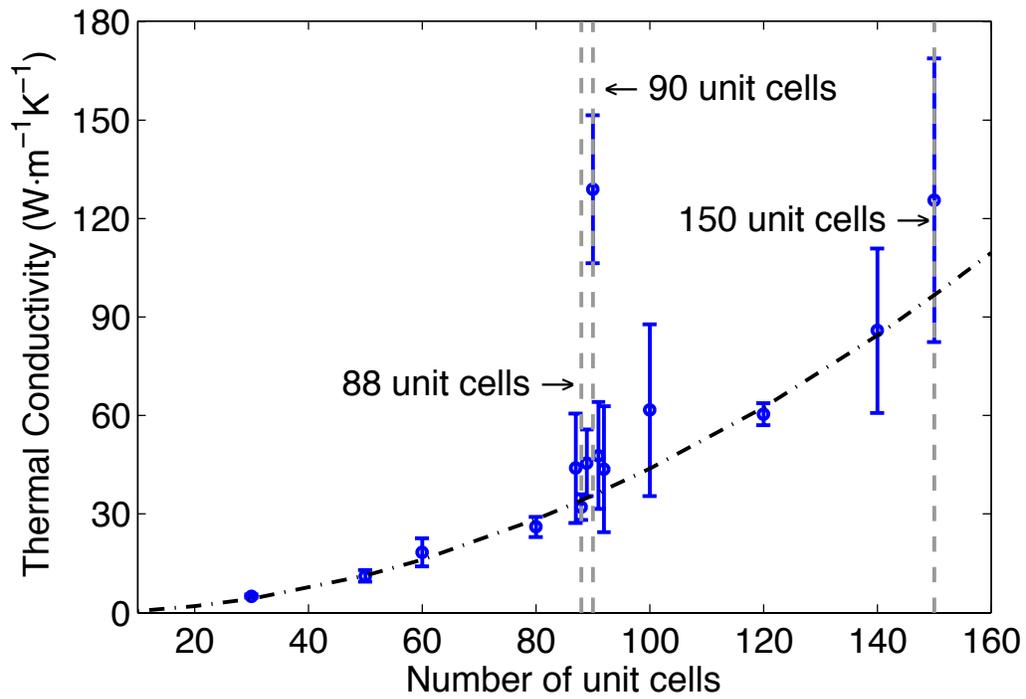

**Figure 2. Finite Pth Chain thermal conductivity versus chain length.** The dash-dot line is a trend line fit to the non-anomalous data.

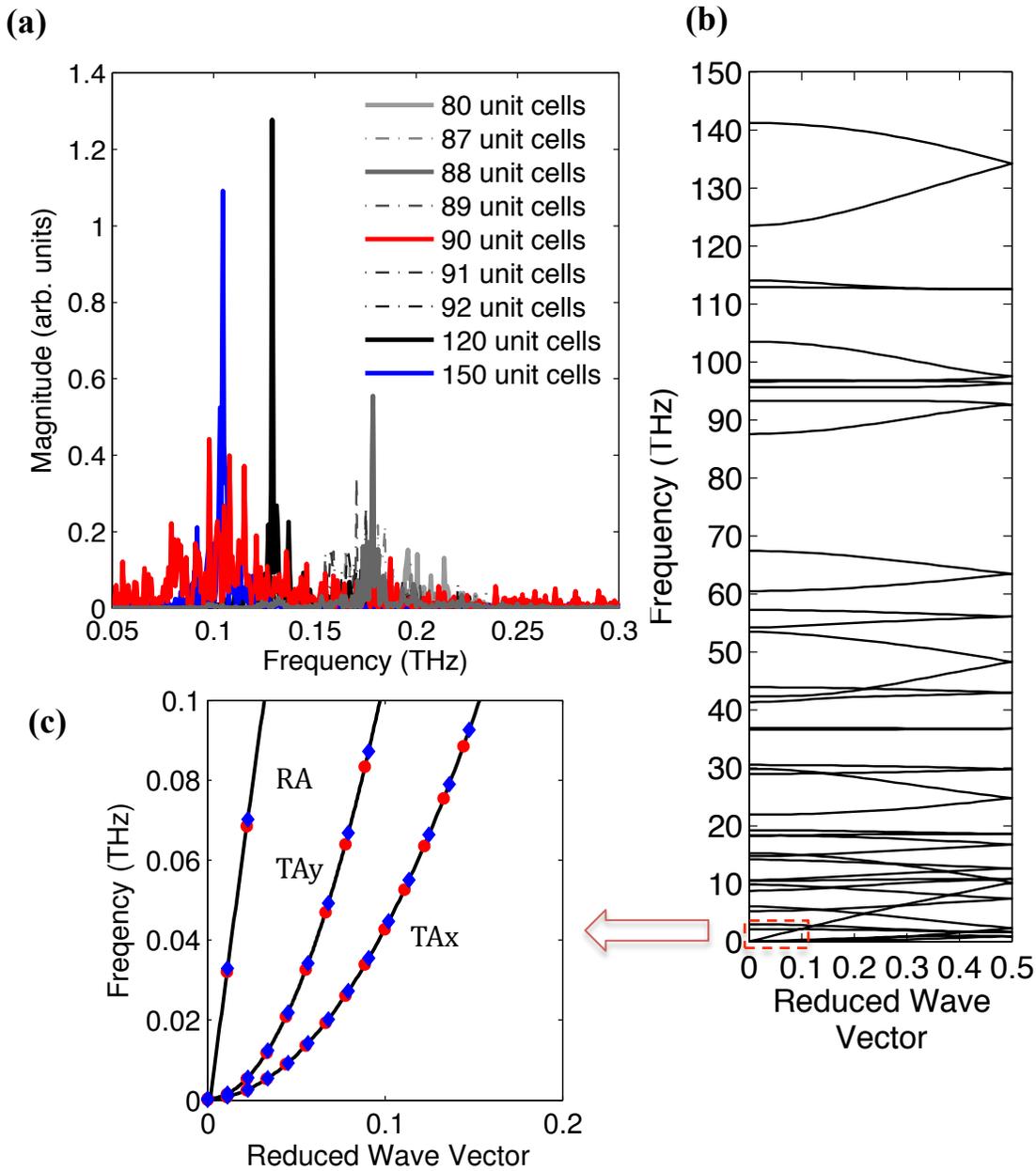

**Figure 3.** (**a**) Power spectrum of the HFAC of different length Pth chains, from the finite chain simulations. (**b**) Phonon dispersion curves for a single Pth chain. (**c**) Zoomed in close up view of the dispersion curves near 0.05 THz, only 3 acoustic branches exist in this range. The discrete modes that exist in the 90 ucs (red circles) and 88 ucs (blue diamonds) modes indicate that the modes are very similar in both cases, even though the HFAC power spectra differ significantly.

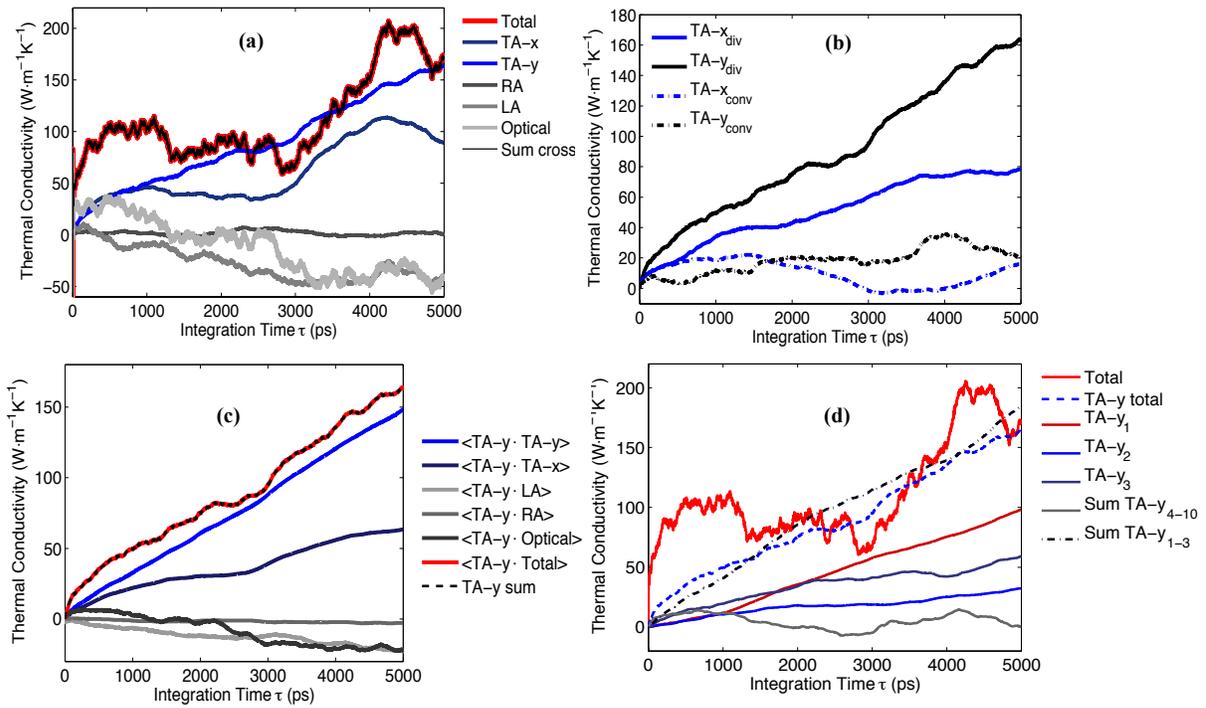

**Figure 4. Thermal conductivity contributions from different polarizations, TA-y branch and TA-y$_1$ mode thermal conductivity contributions.** (**a**) Pth thermal conductivity contribution from different branches. (**b**) TA-x, TA-y branch thermal conductivity contributions in convergent and divergent cases, averaged over 30 ensembles. (**c**) TA-y cross-correlations with other branches. (**d**) TA-y thermal conductivity from each mode on the branch.

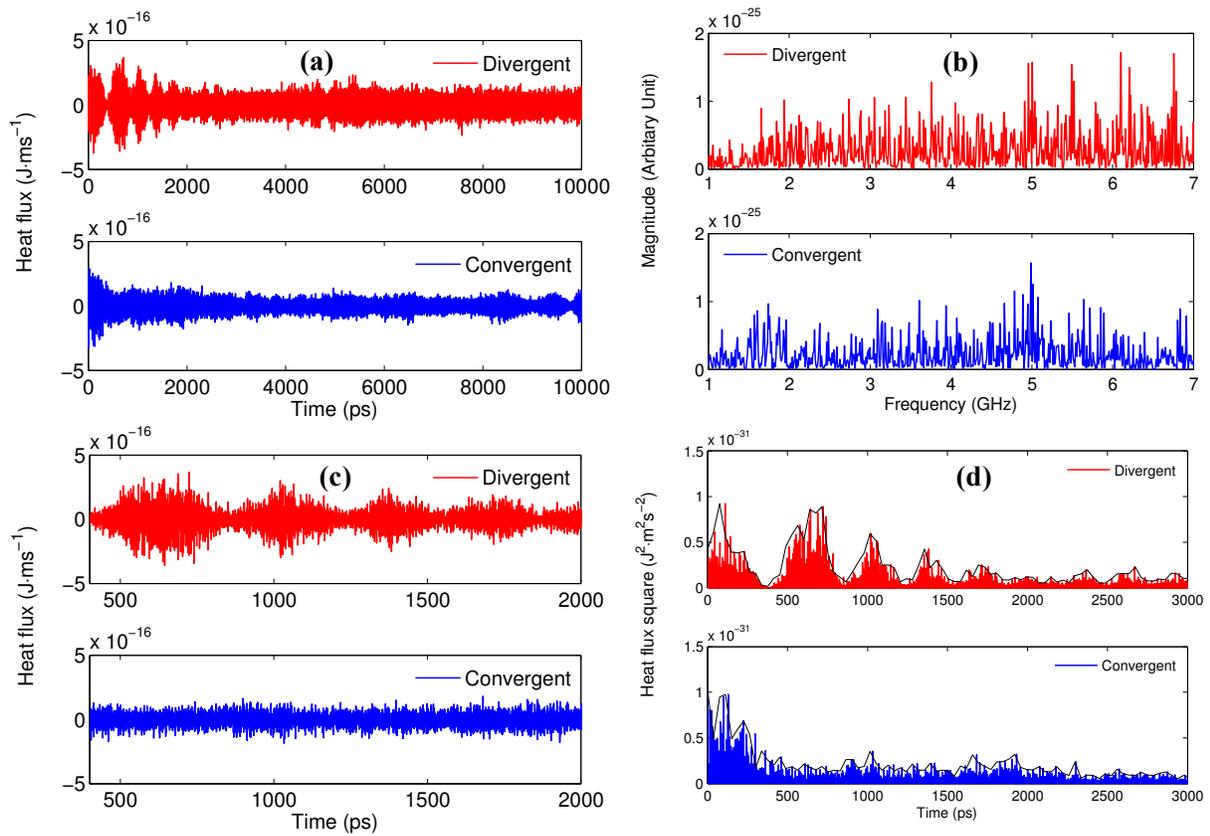

**Figure 5. (a)** TA-y1 mode heat flux for 20 nanoseconds in a divergent and convergent case. **(b)** TA-y1 mode heat flux power spectra for 20-nanosecond length simulation at low frequency (1-7 GHz) in a divergent and convergent case. **(c)** TA-y1 mode heat flux zoomed in from 400 ps to 2000 ps in a divergent and convergent case. **(d)** Square of mode heat flux and the envelop function (black curve) in a divergent and convergent case.

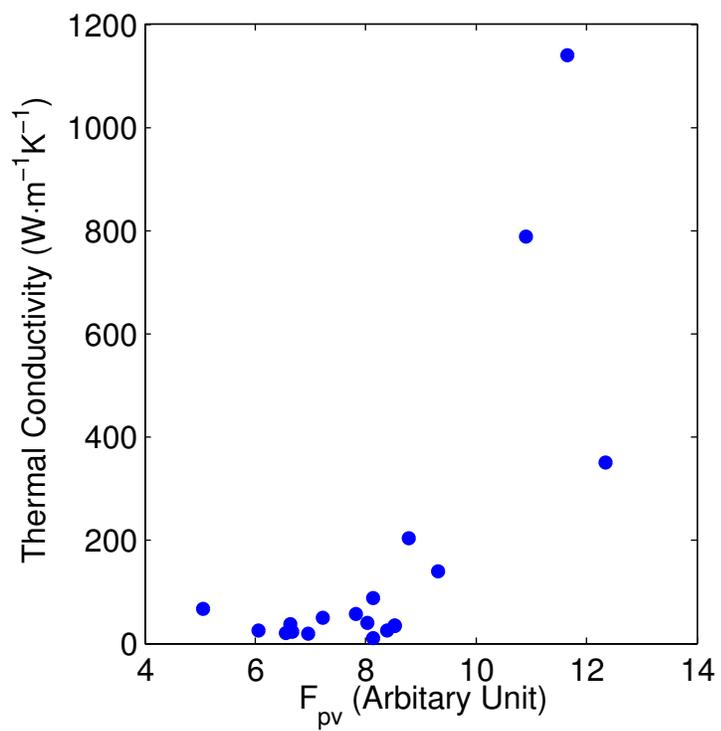

Figure 6. Total thermal conductivity vs. sum of the peak height ($F_{pv}$) as shown in Eq. 7.